\documentclass[sigplan]{acmart}
\setcopyright{none}
\renewcommand\footnotetextcopyrightpermission[1]{}

\usepackage{amsmath,amssymb,amsfonts}
\usepackage{algorithmic}
\usepackage{graphicx}
\usepackage{textcomp}
\usepackage{xcolor}
\usepackage{hyperref}
\usepackage{breakurl}
\usepackage{listings}
\usepackage{footnotebackref}
\usepackage{cleveref}
\usepackage{booktabs}
\usepackage{subcaption}
\usepackage{courier}
\usepackage{array}
\usepackage{verbatim}
\usepackage{caption}
\usepackage{cancel}
\usepackage{multirow}
\usepackage{stmaryrd}
\usepackage{enumitem}
\usepackage{siunitx}
\usepackage{lipsum}
\usepackage[normalem]{ulem}
\useunder{\uline}{\ul}{}
\usepackage{tabularx}
\usepackage{float}
\usepackage{balance}
\PassOptionsToPackage{breaklinks}{hyperref}
\PassOptionsToPackage{linktocpage}{hyperref}

\begin{document}

\newcommand{\todo}[1]{{\textcolor{red}{\textbf{TODO}: #1}}}

\newcommand{\aroma}[0]{\textit{Aroma}}
\newcommand{\codenn}[0]{\textit{CODEnn}}
\newcommand{\cooc}[0]{\textit{COOC}}
\newcommand{\eu}[0]{\textit{UNIF}}
\newcommand{\githubmodel}[0]{\textit{SCS}}
\newcommand{\ncs}[0]{NCS}
\newcommand{\ncspostrank}[0]{NCS\textsubscript{postrank}}
\newcommand{\unifandroid}[0]{UNIF\textsubscript{android}}
\newcommand{\unifstackoverflow}[0]{UNIF\textsubscript{stackoverflow}}

\crefname{lstlisting}{listing}{listings}

\title{Neural Code Search Evaluation Dataset}

\author[Li]{Hongyu Li}
\affiliation{
  \institution{Facebook, Inc.}
  \country{U.S.A}
}
\email{hongyul@fb.com}

\author[Kim]{Seohyun Kim}
\affiliation{
  \institution{Facebook, Inc.}
  \country{U.S.A}
}
\email{skim131@fb.com}

\author[Chandra]{Satish Chandra}
\affiliation{
  \institution{Facebook, Inc.}
  \country{U.S.A}
}
\email{satch@fb.com}

\begin{abstract}
There has been an increase of interest in code search using natural language. Assessing the performance of such code search models can be difficult without a readily available evaluation suite. In this paper, we present an evaluation dataset consisting of natural language query and code snippet pairs, with the hope that future work in this area can use this dataset as a common benchmark. We also provide the results of two code search models (\cite{ncsmapl} and \cite{unifarxiv}) from recent work as a benchmark.
\end{abstract}

\maketitle

\section{Introduction}\label{sec:intro}

In recent years, learning the mapping between natural language and code snippets has been a popular field of research. In particular, \cite{ncsmapl}, \cite{unifarxiv}, \cite{deepcodesearch} have explored finding relevant code snippets given a natural language query, with the models varying from using word embeddings and IR techniques to using sophisticated neural networks. To evaluate the performance of these models, Stack Overflow questions and code answer pairs are prime candidates, as Stack Overflow questions well resemble what a developer may ask. Such an example is "Close/hide the Android Soft Keyboard".\footnote{\url{https://stackoverflow.com/questions/1109022/close-hide-the-android-soft-keyboard}}\footnote{Author: \href{https://stackoverflow.com/users/133858/vidar-vestnes}{Vidar Vestnes}.\\ \url{https://stackoverflow.com/users/133858/vidar-vestnes}} One of the first answers\footnote{\url{https://stackoverflow.com/a/1109108}}\footnote{Author: \href{https://stackoverflow.com/users/822/reto-meier}{Reto Meier}. \url{https://stackoverflow.com/users/822/reto-meier}} on Stack Overflow correctly answers this question. However, collecting these questions can be tedious, and systematically comparing various models can pose a challenge.

To this end, we have constructed an evaluation dataset, which contains natural language queries and relevant code snippet answers from Stack Overflow. It also includes code snippet examples from the search corpus (public repositories from GitHub) that correctly answers each query. We hope that this dataset can be served as a benchmark to evaluate performance across various code search models.

The paper is organized as follows. First we will explain what data we are releasing in the dataset. Then we will describe the process for obtaining this dataset. Finally, we will evaluate two code search models of our own creation, NCS and UNIF, on the evaluation dataset as a benchmark.

\lstset{
    basicstyle=\tiny\ttfamily, 
    tabsize=1,                       
    extendedchars=true,
    breaklines=true,
    frame=single,
    showspaces=false,
    showtabs=false,
    showstringspaces=false,
    escapeinside={<@}{@>}
}

\colorlet{punct}{red!60!black}
\definecolor{background}{HTML}{F8F8F8}
\definecolor{file}{HTML}{EEEEEE}
\definecolor{delim}{RGB}{20,105,176}
\colorlet{numb}{magenta!60!black}
\definecolor{eclipseStrings}{RGB}{42,0.0,255}
\definecolor{eclipseKeywords}{RGB}{127,0,85}

\lstdefinelanguage{json}{
    basicstyle=\scriptsize\ttfamily,
    showstringspaces=false,
    breaklines=true,
    frame=lines,
    backgroundcolor=\color{background},
}

\lstdefinelanguage{filepath}{
    basicstyle=\footnotesize\ttfamily,
    showstringspaces=false,
    breaklines=true,
    frame=none,
    backgroundcolor=\color{background},
}

\section{Dataset Contents}\label{sec:core}
In this section, we explain what data we are releasing.

\subsection{GitHub Repositories}
The most popular Android repositories on GitHub (ranked by the number of stars) is used to create the search corpus. For each repository that we indexed, we provide the link, specific to the commit that was used.\footnote{From August 2018} In total, there are 24,549 repositories.\footnote{There were originally 26,109 repositories - the difference is due to reasons outside of our control (e.g. repositories getting deleted). Note that not all of the links in this dataset may not always be available in the future for the similar reasons.} We will release a text file containing the download links for these GitHub repositories. See \Cref{lst:download-links} for an example.

\begin{figure*}
\begin{lstlisting}[
  language=json,
  escapechar=|,
  caption={GitHub repositories download links example.},
  captionpos=b,
  label={lst:download-links}
]
https://github.com/00-00-00/ably-chat/archive/9bb2e36acc24f1cd684ef5d1b98d837055ba9cc8.zip
https://github.com/01sadra/Detoxiom/archive/c3fffd36989b0cd93bd09cbaa35123b9d605f989.zip
https://github.com/0411ameya/MPG_update/archive/27ac5531ca2c2f123e0cb854ebcb4d0441e2bc98.zip
https://github.com/0508994/MinesweeperGO/archive/ba0e0e45d2da21dde2365ce09277aad511de6885.zip
https://github.com/07101994/My-PPT-Presentation/archive/b89b17a962d5c3e5682fa751228a9f9ca593d77b.zip
https://github.com/0912718/ICT-lab/archive/d1d723edb722013cc83761f0f9df252cfd3361c3.zip
https://github.com/0Cubed/ZeroMediaPlayer/archive/d84c675f9dc8b16f823bb252db9ee368fbd5cd8e.zip
...
\end{lstlisting}
\end{figure*}

\subsection{Search Corpus}
The search corpus is indexed using all method bodies parsed from the 24,549 GitHub repositories. In total, there are 4,716,814 methods in this corpus. The code search model will find relevant code snippets (i.e. method bodies) from this corpus given a natural language query. In this data release, we will provide the following information for each method in the corpus:
\begin{itemize}[leftmargin=7mm]
  \small
  \item \textbf{id:} Each method in the corpus has a unique numeric identifier. This ID number will also be referenced in our evaluation dataset.
  \item \textbf{filepath:} The file path is in the format of
\begin{lstlisting}[language=filepath, escapechar=|]
|:owner/:repo/relative-file-path-to-the-repo|
\end{lstlisting}
  \item \textbf{method\_name}
  \item \textbf{start\_line:} Starting line number of the method in the file.
  \item \textbf{end\_line:} Ending line number of the method in the file.
  \item \textbf{url:} GitHub link to the method body with commit ID and line numbers encoded.
\end{itemize}
\Cref{lst:search-corpus} provides an example of a method in the search corpus.

\begin{lstlisting}[
  float,
  language=json,
  escapechar=|,
  caption={Search corpus example.},
  captionpos=b,
  label={lst:search-corpus}
]
 |\textbf{\{}|
   |\textbf{\textcolor{delim}{"id"}}|: 4716813,
   |\textbf{\textcolor{delim}{"filepath"}}|: "Mindgames/VideoStreamServer/playersdk/src/main/java/com/kaltura/playersdk/PlayerViewController.java",
   |\textbf{\textcolor{delim}{"method\_name"}}|: "notifyKPlayerEvent",
   |\textbf{\textcolor{delim}{"start\_line"}}|: 506,
   |\textbf{\textcolor{delim}{"end\_line"}}|: 566,
   |\textbf{\textcolor{delim}{"url"}}|: "https://github.com/Mindgames/VideoStreamServer/blob/b7c73d2bcd296b3a24f83cf67d6a5998c7a1af6b/playersdk/src/main/java/com/kaltura/playersdk/PlayerViewController.java\#L506-L566"
 |\textbf{\}}|
\end{lstlisting}

\subsection{Evaluation Dataset}
The evaluation dataset is composed of 287 Stack Overflow question and answer pairs, for which we release the following information:
\begin{itemize}[leftmargin=7mm]
  \small
  \item \textbf{stackoverflow\_id:} Stack Overflow post ID.
  \item \textbf{question:} Title of the Stack Overflow post.
  \item \textbf{question\_url:} URL of the Stack Overflow post.
  \item \textbf{answer:} Code snippet answer to the question.
  \item \textbf{answer\_url:} URL of the Stack Overflow answer to the question.
  \item \textbf{examples:} 3 methods from the search corpus that best answer the question (most similar to the Stack Overflow answer).
  \item \textbf{examples\_url:} GitHub links to the examples.
\end{itemize}

Note that there may be more acceptable answers to each question. See \Cref{lst:evaluation-dataset} for a concrete example of an evaluation question in this dataset. The source of the question and answer pairs is extracted from the Stack Exchange Network \cite{stackexchange}.

\begin{lstlisting}[
  float,
  language=json,
  escapechar=|,
  caption={Evaluation dataset example.},
  captionpos=b,
  label={lst:evaluation-dataset}
]
 |\textbf{\{}|
   |\textbf{\textcolor{delim}{"stackoverflow\_id"}}|: 1109022,
   |\textbf{\textcolor{delim}{"question"}}|: "Close/hide the Android Soft Keyboard",
   |\textbf{\textcolor{delim}{"question\_url"}}|: "https://stackoverflow.com/questions/1109022/close-hide-the-android-soft-keyboard",
   |\textbf{\textcolor{delim}{"question\_author"}}|: "Vidar Vestnes",
   |\textbf{\textcolor{delim}{"question\_author\_url"}}|:
     "https://stackoverflow.com/users/133858",
   |\textbf{\textcolor{delim}{"answer"}}|: "// Check if no view has focus:\nView view = this.getCurrentFocus();\nif (view != null) {   InputMethodManager imm = (InputMethodManager)getSystemService(Context.INPUT_METHOD_SERVICE);  imm.hideSoftInputFromWindow(view.getWindowToken(), 0);}",
   |\textbf{\textcolor{delim}{"answer\_url"}}|: "https://stackoverflow.com/a/1109108",
   |\textbf{\textcolor{delim}{"answer\_author"}}|: "Reto Meier",
   |\textbf{\textcolor{delim}{"answer\_author\_url"}}|:
     "https://stackoverflow.com/users/822",
   |\textbf{\textcolor{delim}{"examples"}}|: |\textbf{[}|1841045, 1800067, 1271795|\textbf{]}|,
   |\textbf{\textcolor{delim}{"examples\_url"}}|: |\textbf{[}|
     "https://github.com/alextselegidis/easyappointments-android-client/blob/39f1e8...",
     "https://github.com/zelloptt/zello-android-client-sdk/blob/87b45b6...",
     "https://github.com/systers/conference-android/blob/a67982abf54e0...",
   |\textbf{]}|
 |\textbf{\}}|
\end{lstlisting}

\subsection{NCS / UNIF Score Sheet}
We provide the evaluation results for two code search models of our creation, each with two variations:
\begin{itemize}[leftmargin=7mm]
  \item \ncs{}: an unsupervised model which uses word embedding derived directly from the search corpus\cite{ncsmapl}.
  \item \ncspostrank{}: an extension of the base NCS model that performs a post-pass ranking, as explained in \cite{ncsmapl}.
  \item \unifandroid{}, \unifstackoverflow{}: a supervised extension of the NCS model that uses a bag-of-words-based neural network with attention. The supervision is learned using \textit{GitHub-Android-Train} and \textit{StackOverflow-Android-Train} datasets, respectively, as described in \cite{unifarxiv}.
\end{itemize}
We provide the rank of the first correct answer (FRank) for each question in our evaluation dataset. The score sheet is saved in a comma-delimited csv file as illustrated in \Cref{lst:score-sheet}.

\begin{lstlisting}[
  language=json,
  escapechar=|,
  caption={Score sheet example. "NF" stands for correct answer not found in the top 50 returned results.},
  captionpos=b,
  label={lst:score-sheet}
]
No.,StackOverflow ID,NCS FRank,NCS_postrank FRank,UNIF_android FRank,UNIF_stackoverflow FRank
1,1109022,NF,1,1,1
2,4616095,17,1,31,19
3,3004515,2,1,5,2
4,1560788,1,4,5,1
5,3423754,5,1,22,10
6,1397361,NF,3,2,1
\end{lstlisting}

\section{How we Obtained the Dataset}\label{sec:obtaining}
In this section, we describe the procedure for how we obtained the data.

\textbf{GitHub repositories.}
We obtained the information of the GitHub repositories with the GitHub REST API \cite{githubapi}, and the source files were downloaded using publicly available links.

\textbf{Search corpus.}
The search corpus was obtained by dividing each file in the GitHub repositories by method-level granularity.

\textbf{Evaluation dataset.}
The benchmark questions were collected from a data dump publicly released by Stack Exchange \cite{stackexchange}. To select the set of Stack Overflow question and answer pairs, we created a heuristics-based filtering pipeline where we discarded open-ended, discussion-style questions. We first obtained the most popular 17,000 questions on Stack Overflow with ``Android'' and ``Java'' tags. The dataset is further filtered with the following criteria: 1) there exists an upvoted code answer, 2) the ground truth code snippet has at least one match in the search corpus. From this pipeline, we were able to obtain 518 questions. Finally, we manually went through these questions and filtered out questions with vague queries and/or code answers. The final dataset contains 287 Stack Overflow question and answers pairs.

\textbf{NCS / UNIF score sheet.}
To judge whether a method body correctly answers the query, we compare how similar it is to the Stack Overflow answer - we do this systematically using a code-to-code similarity tool, called Aroma \cite{aromaarxiv}. Aroma gives a similarity score between two code snippets; if this score is above a certain threshold (0.25 in our case), we count it as success. This similarity score, aims to mimic manually assessing the correctness of search results in an automatic and reproducible fashion, while leaving out human judgment in the process. More details on how we chose this threshold can be found in \cite{unifarxiv}.


\section{Evaluation}\label{sec:evaluation}
We provide the results for four models: \ncs{}, \ncspostrank{}, \unifandroid{}, and \unifstackoverflow{}.

\Cref{tab:results} reports the number of questions answered within the top\_n returned code snippet, where n = 1, 5, and 10 (Answered@1, 5, 10 in \Cref{tab:results}), as well as the Mean Reciprocal Rank (MRR).

\begin{table}
\caption{Number of questions answered in the top 1, 5, 10 and MRR for \ncs{}, \ncspostrank{}, \unifandroid{} and \unifstackoverflow{}.}
\label{tab:results}
\centering
\scriptsize
\begin{tabular}{ccccc}
\toprule
Model        & Answered@1 & Answered@5 & Answered@10 & MRR\\ \midrule
\ncs{} & 33 & 74 & 98  & 0.189 \\
\ncspostrank{} & 85 & 151 & 180 & 0.4  \\
\unifandroid{}  & 25    & 74 & 110 & 0.178   \\
\unifstackoverflow{} & \textbf{104} & \textbf{164} & \textbf{188} & \textbf{0.465} \\
\bottomrule
\end{tabular}
\end{table}

\balance
\bibliographystyle{plainurl}
\bibliography{references}

\end{document}